\documentclass[aps,pra,reprint,showpacs,amssymb,amsmath]{revtex4-1}
\usepackage{graphicx}
\def\be{\begin{equation}}
\def\ee{\end{equation}}
\def\ber{\begin{eqnarray}}
\def\eer{\end{eqnarray}}
\def\bern{\begin{eqnarray*}}
\def\eern{\end{eqnarray*}}

\def\rv{\mathbf{r}}

\def\Gv{\mathbf{G}}

\def\pv{\mathbf{p}}
\def\qv{\mathbf{q}}

\def\0v{\mathbf{0}}
\def\1v{\mathbf{1}}
\def\2v{\mathbf{2}}
\def\3v{\mathbf{3}}

\begin{document}

\title{Electron energy-loss spectroscopy of quasi-two-dimensional crystals: Beyond the energy-loss functions formalism}
\author{Vladimir~U.~Nazarov}
\affiliation{Research Center for Applied Sciences, Academia Sinica, Taipei 11529, Taiwan}
\email{nazarov@gate.sinica.edu.tw}

\author{Vyacheslav M. Silkin}
\affiliation{Departamento de F\'{i}sica de Materiales, Facultad de Ciencias Qu\'{i}imicas, Universidad del Pais Vasco/Euskal Herriko Unibertsitatea, Apdo. 1072, San Sebasti\'{a}n/Donostia, 20080 Basque Country, Spain}
\affiliation{Donostia International Physics Center (DIPC), Paseo Manuel de Lardizabal 4, San Sebasti\'{a}n/Donostia, 20018 Basque Country, Spain}
\affiliation{IKERBASQUE, Basque Foundation for Science, 48013 Bilbao, Spain}

\author{Eugene E. Krasovskii}
\affiliation{Departamento de F\'{i}sica de Materiales, Facultad de Ciencias Qu\'{i}imicas, Universidad del Pais Vasco/Euskal Herriko Unibertsitatea, Apdo. 1072, San Sebasti\'{a}n/Donostia, 20080 Basque Country, Spain}
\affiliation{Donostia International Physics Center (DIPC), Paseo Manuel de Lardizabal 4, San Sebasti\'{a}n/Donostia, 20018 Basque Country, Spain}
\affiliation{IKERBASQUE, Basque Foundation for Science, 48013 Bilbao, Spain}

\begin{abstract}
A consistent theory of electron energy-loss spectroscopy (EELS) includes two indispensable elements:
(i) electronic response of the target system  and  (ii)  quantum kinematics of probing electrons.
While for the bulk materials and their surfaces separating these two aspects and focusing on the former 
is the usual  satisfactory practice (the energy-loss functions formalism), we show
that, for quasi-2D crystals, 
the interplay of the system's electronic response and the details of the probe's motion affects EEL spectra  dramatically, 
and it must  be taken into account for the reliable interpretation of the experiment.
To this end, we come up with a unified theory which, on the same footing, treats  both 
the long- and short-range scattering, 
within both the transmission and reflection experimental setups.  
Our calculations performed for graphene reveal a phenomenon of a strong coupling between the $\pi+\sigma$ plasmon excitation
and elastic {\it scattering resonances}.
Freed from the conventions of the  energy-loss functions formalism,
our theory serves as a consistent and systematic means of the understanding  of the EELS of quasi-2D materials.
\end{abstract}

\maketitle

While  electron energy-loss spectroscopy (EELS) is a powerful experimental tool in the  studies of the growing family
of quasi-two-dimensional (Q2D) materials \cite{Egerton-09}, its theoretical support for these systems  remains unsatisfactory, 
being  based on the often irrelevant analogies with  the bulks and  surfaces of the 3D solids \cite{Nazarov-15}. 
For the latter, the  EEL spectra interpretation traditionally
relies on the concept of the energy-loss functions, such as $-\text{Im } 1/\epsilon(\qv,\omega)$ \cite{Pines-66},
where $\epsilon(\qv,\omega)$  is the wave-vector and frequency-dependent dielectric function. 
Another popular energy-loss function, which came from the field of surface science, is $-\text{Im } g(\qv_\|,\omega)$,
where $g(\qv_\|,\omega)$ is the so-called $g$-function \cite{Liebsch},
$\qv_\|$ being the in-plane  wave-vector transferred to the system.

The description in terms of  loss-functions is convenient, 
since the latter are  properties of the target system alone, saving us the trouble of considering details of the EELS experiment. 
Moreover, doing so is justified,
as long as the characteristics of the {\it elastic} scattering at the crystal lattice
change slowly within the energy-range of interest, 
constituting a background to the sharp features of the energy-losses due to the {\it inelastic} scattering
at the electronic sub-system of the target. This condition is usually satisfied
in the EELS of the bulk solids (films) and surfaces, which has led to the loss-functions formalism
becoming generally adopted in the EELS theory, and later, with the emergence of graphene 
\cite{Novoselov-04},
automatically transferred to the field of  Q2D materials.
\begin{figure}[h!]
\includegraphics[width=  \columnwidth, clip=true, trim=50 75 25 65]{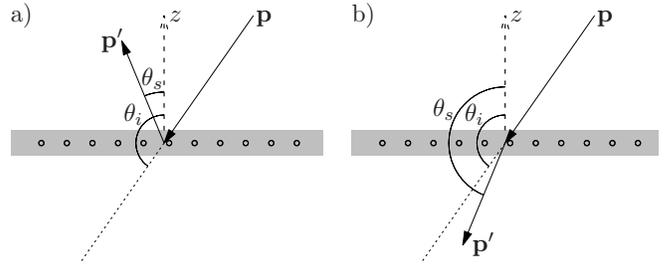}
\caption{\label{geom}
Schematic illustration of the reflection a) and transmission b) geometries of the EELS experiment on a Q2D crystal.}
\end{figure}

Recent advancements in  the understanding of the interaction of electron beams with Q2D crystals have, however, revealed
that the electrons' elastic scattering  at these systems
is far from changing slowly with the energy. 
In particular, sharp peaks and dips in the energy-dependence of the reflection
and transmission coefficients were predicted theoretically in graphene and identified as 
{\it scattering resonances}  in Q2D crystals \cite{Nazarov-13}
(metastable unoccupied states due to the in-plane and the perpendicular
motions being coupled by the periodic potential), which 
has later been confirmed and studied experimentally \cite{Pisarra-14-2,Jobst-15,Wicki-16,Krivenkov-17}.
As a consequence, the reliable interpretation of EEL spectra of Q2D crystals becomes impossible
without a comprehensive theory which takes into account all the facets of the probe--target interactions. 
Such an approach lacking presently, the purpose of this work is to fill the gap 
by constructing the  theory of EELS in application to Q2D crystals
in the natural terms of the quantum-mechanical scattering of  a charge at a many-body system.
On this way, all the features of the inelastic and elastic scattering are included, as well as, importantly, the effects of their intermixture.  

We are concerned with   the EELS experiment on a Q2D crystal as schematized in Fig.~\ref{geom} for
both the reflection and transmission regimes.
The starting point of our approach is the general formula for the differential cross-section of a probe 
electron's scattering  at the electronic sub-system of a target, 
accompanied by the interaction with the target's  lattice potential.
This fomular, in the real-space representation, reads \cite{Nazarov-95S,Nazarov-99,Nazarov-01}
\begin{equation}
\begin{split}
\frac{d ^2\sigma}{d \omega d \Omega} (\pv'\leftarrow \pv) & = 
-\frac{16 \pi^3 p'}{p} {\rm Im} \int \frac{\rho^*(\rv)}{|\rv-\rv_1|}  \\
& \times \chi(\rv_1,\rv_1',\omega) \frac{\rho(\rv')}{|\rv_1'-\rv'|}  d \rv d \rv' 
d\rv_1 d\rv_1'.
\end{split}
\label{DCS}
\end{equation}
In Eq.~(\ref{DCS}),
$\chi$ is the interacting  density-response function of the target system,
$\pv$ and $\pv'$ are the momenta of the probing electron before and after the scattering, respectively,
$\omega=(p^2-p'^2)/2$ is the energy transferred to the target,
\begin{equation}
\rho(\rv)=\langle \rv |\pv^+\rangle^* \times \langle \rv  \,|\pv'^-\rangle
\label{rho}
\end{equation}
is the complex-valued `charge-density' determined   by the elastically scattered waves $|\pv^\pm\rangle$, 
which are solutions of the Lippmann-Schwinger equations \cite{Taylor}
\begin{equation}
|\pv^\pm\rangle=|\pv \rangle+G^0 \left(\frac{p^2}{2 }\pm i 0_+ \right) V_l |\pv^\pm\rangle,
\label{LS}
\end{equation}
$\langle\rv|\pv \rangle=(2\pi)^{-3/2} e^{i\pv\cdot\rv}$ is plane-wave,
$V_l(\rv)$ is the lattice potential,
$G^0(E)=(E-\hat{H}_0)^{-1}$ and $\hat{H}_0=-\frac{1}{2} \Delta + V_l(\rv)$ are the non-interacting Green's function and the 
Hamiltonian, respectively, and $0_+$ is an infinitesimal positive.
Equations (\ref{DCS})-(\ref{LS}) solve the inelastic scattering  problem at an arbitrary (many- or few-) electron
system with the interaction between the probe charge
and the electronic sub-system of the target
accounted for in the first Born approximation, while the probe--lattice interaction is included to all orders \cite{Nazarov-95S}
(distorted-wave approximation \cite{Taylor}). We  note that  Eqs.~(\ref{DCS})-(\ref{LS}) 
include the long- and short range (dipole- and impact \cite{Liebsch}, respectively) scattering regimes as the two specific cases.

To make connection to the loss-function formalism, we note that, if the elastic scattering is neglected, 
which means that only the first term in the right-hand side of Eq.~(\ref{LS}) 
is kept, Eq.~(\ref{rho}) gives  $\rho(\rv)=e^{i(\pv'-\pv)\cdot \rv}/(2\pi)^3$, and Eq.~(\ref{DCS}) integrates to
\begin{equation}
\frac{d ^2\sigma}{d \omega d \Omega} (\pv'\leftarrow \pv)  = 
-\frac{32\pi^2 p'}{ |\Delta \pv|^4 p} {\rm Im} \
\chi(\Delta\pv,\Delta \pv,\omega),
\label{bulk}
\end{equation}
where $\Delta \pv=\pv-\pv'$, and $\chi$ is written  in the reciprocal space representation. 
If the target is a bulk solid, then
the usual bulk energy-loss function $- {\rm Im}\ 1/\epsilon(\Delta \pv,\omega)$
is readily retrieved from the right-hand side of  Eq.~(\ref{bulk}). 
On the other hand, for a Q2D crystal, Eq.~(\ref{bulk}) coincides (to a coefficient) with the loss function of Ref.~\cite{Nazarov-15}
in the transmission geometry
(for the connection to the $g$--function of the surface scattering, see Appendix \ref{AA}).

Returning to the  the simultaneous inelastic and elastic scattering, 
we note that, in a Q2D crystal, the in-plane component of the wave-vector conserves to within a reciprocal lattice vector $\Gv$.
As a consequence, the density-response function becomes a matrix in the reciprocal lattice vectors
$\chi_{\Gv \Gv'}(z,z',\qv,\omega)$, where the in-plane wave-vector $\qv$ belongs to the first Brillouin zone.
Equation (\ref{DCS}) is then conveniently  transformed to
\begin{widetext}
\begin{equation}
\begin{split}
\frac{1}{A} \frac{d \sigma}{d \omega d \Omega} (\pv'\leftarrow \pv) &= 
-\frac{64 \pi^5 p'}{p} {\rm Im} 
\sum\limits_{\substack{\Gv \tilde{\Gv}\\ \Gv' \tilde{\Gv}'}}
\int \chi_{\Gv \Gv'}(z_1,z_1',\qv,\omega) 
\frac{e^{-|\Gv+\qv| |z_1-z|} e^{-|\Gv'+\qv| |z_1'-z'|}  }{|\Gv+\qv| |\Gv'+\qv|}  \\
& \times a^{+*}_{\pv,\Gv+\tilde{\Gv}}(z) a^-_{\pv',\tilde{\Gv}+\Gv_0}(z)
a^+_{\pv,\Gv'+\tilde{\Gv}'}(z') a^{-*}_{\pv',\tilde{\Gv}'+\Gv_0}(z')  
 d z d z' d z_1 d z_1',
\end{split}
\label{cross2D}
\end{equation}
\end{widetext}
where $a^\pm_{\pv,\Gv}(z)$ are the Fourier coefficients in the expansion
\begin{equation}
\langle \rv |\pv^\pm\rangle= \sum\limits_{\Gv} a^\pm_{\pv,\Gv}(z) e^{i(\Gv+\pv_\|)\cdot\rv_\|},
\label{apm}
\end{equation}
$A$ is the normalization area, and $\Gv_0$ reduces the parallel component of the transferred momentum
to the first Brillouin zone: $\pv_\|-\pv'_\|=\qv+\Gv_0$.

The practical implementation of the approach based on Eq.~(\ref{cross2D})  includes the following major steps:
\begin{enumerate}
\item
\label{A}
Calculation of the interacting  density-response function $\chi$ of the Q2D crystal. This is done within the framework
of the time-dependent density-functional theory (TDDFT) with the use of the relation \cite{Gross-85} 
\begin{equation}
\chi^{-1}=\chi_s^{-1}-f_H-f_{xc},
\label{TDDFT}
\end{equation}
where $\chi_s$ is the Kohn-Sham (KS) \cite{Kohn-65} independent electrons density-response function,
and $f_H$ and $f_{xc}$ are the Hartree and the exchange-correlation kernels of TDDFT, respectively \cite{Gross-85}.
\item
\label{B}
Calculation of  the elastic scattering wave-functions $|\pv^\pm\rangle$ of the probe electron by solving Eqs.~(\ref{LS}) for the Q2D crystal.
In other words, the $z$-dependent coefficients $a^\pm_{\pv,\Gv}(z)$ in Eq.~(\ref{apm}) must be found.
\item
Since the super-cell method (substituting the Q2D crystal with an infinite periodic array of such crystals) is used for a practicable solution of
the sub-problems \ref{A} and \ref{B}, the construction of the quantities pertinent to the single Q2D system from those of the array 
of such systems
is of major importance. For the density-response function $\chi$, we use the method of the elimination of the spurious interaction between
the fictitious copies of the Q2D crystal \cite{Nazarov-15}. To find the elastic scattering wave-functions, we first solve the band-structure
problem of the infinite array system, then, from it, we construct $|\pv^\pm\rangle$ of the single Q2D crystal by imposing the proper asymptotic conditions in vacuum (see Appendix \ref{AB}). These conditions 
ensure $|\pv^+\rangle$ and $|\pv^-\rangle$ to be the low-energy electron diffraction (LEED) 
and  the time-reversed LEED wave-functions, respectively,  of the Q2D crystal \cite{Nazarov-16}.  
\end{enumerate}

Our calculations use the local-density approximation (LDA) for the ground-state KS problem \cite{Kohn-65}
and the random-phase approximation (RPA) [setting $f_{xc}=0$ in Eq.~(\ref{TDDFT})] for the dynamic response.

\begin{figure}[h!]
\includegraphics[width=  \columnwidth, clip=true, trim=65 0 50 0]{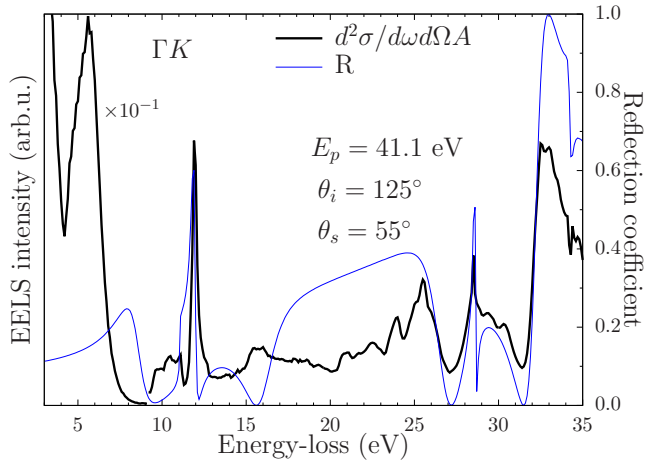}
\caption{\label{sig}
Calculated EEL reflection spectrum of  graphene (thick black line) and the coefficient of reflection (thin blue line plotted against the right $y$-axis). 
For better visualization, the spectrum is split into two parts, of which the  low-energy one is scaled by 0.1.
}
\end{figure}
First, we present results  corresponding to the experimental setup of the high-resolution EELS (HREELS) \cite{Ibach-82}.
In Fig.~\ref{sig}, the reflection EEL spectrum of graphene calculated with the use of the present theory is plotted together
with the reflection coefficient (the latter changing with $\pv'$, while $\pv$ is fixed).
The energy of the incident electrons is $E_p=41.1$ eV, the angle of incidence is $\theta_i=125^\circ$,
and the angle of scattering is $\theta_s=55^\circ$ (polar angles are counted relative to the positive $z$-axis,
see Fig.~\ref{geom}).
A strong peak of the $\pi$--plasmon ($\sim5.6$ eV at our geometry) is little affected by the elastic scattering.
It is, however, instructive to see how the influence of the elastic scattering changes the spectrum in the energy range of the $\pi+\sigma$ plasmon,
the latter extending broadly from about $10$ eV  to $25$ eV within the energy-loss function approach \cite{Nazarov-15}.
A sharp peak in the reflection coefficient due to the scattering resonance \cite{Nazarov-13} at $\sim11.9$ eV
leads to a  peak in the EELS intensity at this energy. 
The same happens at $\sim 25.5$, $28.5$, and $33$ eV.
Generally, the EEL spectrum becomes a product of the interplay
of the inelastic and elastic processes. At the same time, it would be an oversimplification to conclude that the EEL spectrum
just follows the reflectance one: In Eq.~(\ref{cross2D}) the reflectance coefficient does not factorize and, therefore,
the influence of the elastic scattering on the EELS is not  straightforward. 
This can be observed in Fig.~\ref{sig}, considering a maximum in the EEL spectrum at $\sim 15.7$ eV, where reflectance has a minimum.
\begin{figure}[h!]
\includegraphics[width=  \columnwidth, clip=true, trim=65 0 50 0]{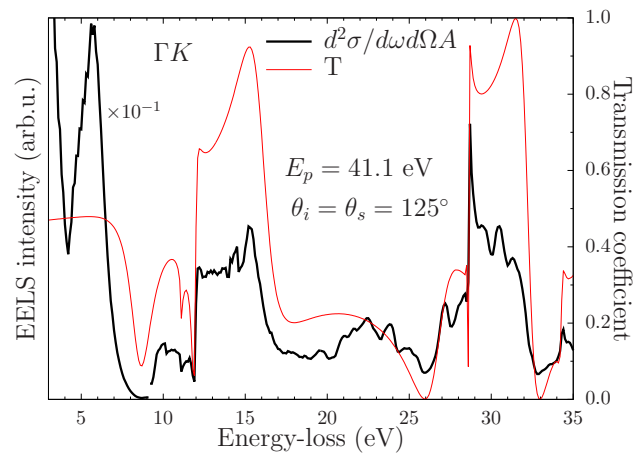}
\caption{\label{sigt}
Calculated EEL transmission spectrum of  graphene (thick black line) and the coefficient of transmission (thin red line plotted against the right $y$-axis). 
For better visualization, the spectrum is split into two parts, of which the  low-energy one is scaled by 0.1.
}
\end{figure}

Similar observations can be made from the EELS  in the transmission geometry in comparison with the transmission 
coefficient, as presented in Fig.~\ref{sigt}. We conclude that the elastic scattering
affects the EEL spectrum dramatically, especially so in the region of the $\pi+\sigma$ plasmon. Different parts of the spectrum are 
strongly enhanced and suppressed in the reflection and transmission regimes, which is due to the presence of the scattering
resonances in this energy range.

HREEL experimental spectra of the free-standing graphene are not, to the best of our knowledge, available in the literature so far. 
Although measurements on graphene supported on
substrates have been reported \cite{Lu-09,Politano-11,Cupolillo-15}, 
the presence of a substrate can affect both the scattering resonances and the electronic response,
making impossible the quantitative comparison with the theory for the free-standing graphene. On the other hand,
inclusion of a substrate in the {\it ab initio}  theory is a very challenging task, remaining a matter of the future.
For the discussion in conjunction with  experiment we, therefore,
turn to the EELS in the transmission electron microscope (TEM) \cite{Egerton-09}.

EELS measurements on free-standing 2D materials   are  conducted in TEM 
using energetic ($\sim 40 - 120$ keV) incident electron beams \cite{Eberlein-08,Wachsmuth-13,Liou-15}. 
For energies that high,
it is practically impossible to obtain $|\pv^\pm\rangle$ of Eq.~(\ref{LS})
from the band-structure calculation, but, fortunately, this is also unnecessary, since, in this case, the first Born approximation
should already 
provide an accurate solution to the elastic scattering problem.
We, therefore, use  Eq.~(\ref{cross2D}) again, but with the coefficients $a^\pm_{\pv,\Gv}(z)$  found to the first order in the
magnitude of $V_l(\rv)$  (see Appendix \ref{AC}). 

\begin{figure}[h!]
\includegraphics[width=  \columnwidth, clip=true, trim=60 0 10 0]{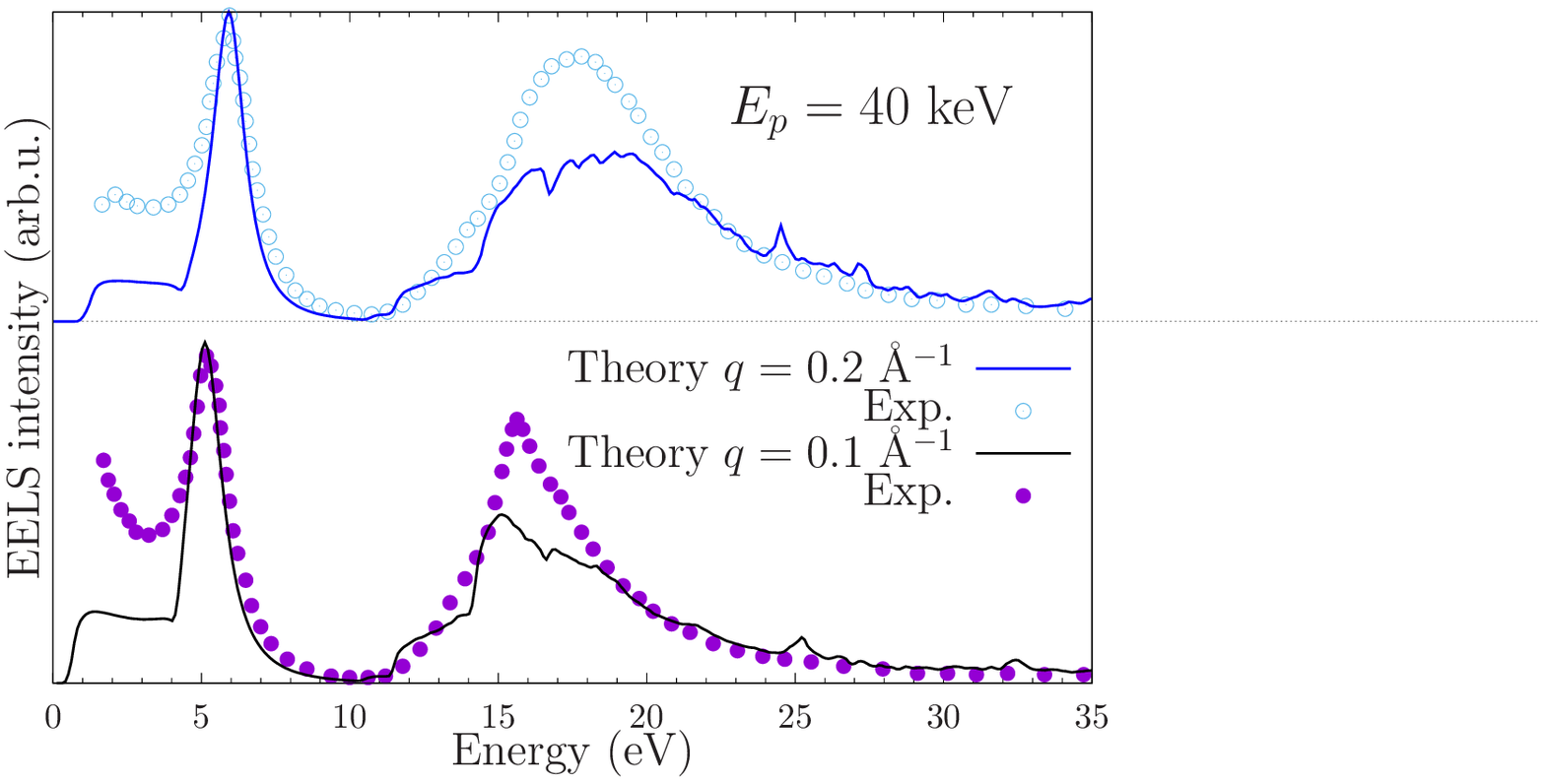}
\caption{\label{Wachsmuth}
Calculated EEL spectrum of graphene
(solid lines) and the experimental EELS in TEM (circles). Experimental data are digitized
from Ref.~\cite{Wachsmuth-13}. Spectra are normalized to the $\pi$-plasmon amplitude.
}
\end{figure}

Scattering resonances do not exist or are negligible in the keV energy-range, and the influence of the elastic scattering
on the inelastic one  differs for EELS in TEM from that for HREELS, while, as we show below, it still remains important. 
Results of our calculations  presented in Fig.~\ref{Wachsmuth} correspond to the setup and are compared to the experiment of Ref.~\cite{Wachsmuth-13}.
The geometry of this experiment suggests that the change in the in-plane momentum of the projectile 
$\Delta \pv_\|$ 
belongs to the first Brillouin zone of graphene. 
As a consequence (in the absence of scattering resonances), the full calculation with the use of Eq.~(\ref{cross2D})
results in a spectrum indistinguishable from that obtained with Eq.~(\ref{bulk}) for the energy-loss function (not shown).

\begin{figure}[h!]
\includegraphics[width=  \columnwidth, clip=true, trim=60 0 10 0]{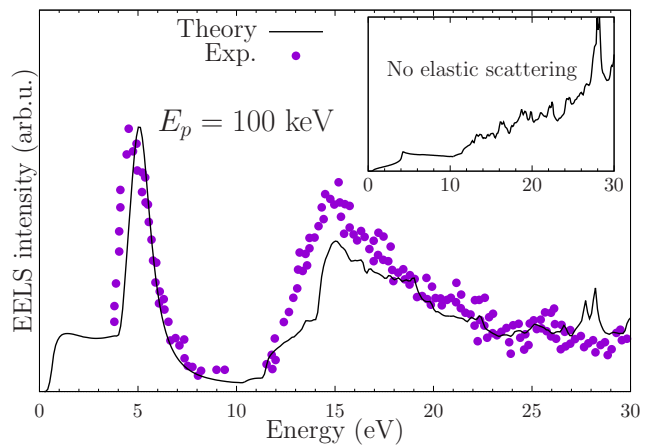}
\caption{\label{Eberlein}
Calculated EEL spectrum of graphene
(solid line) and the experimental EELS in TEM (circles). Experimental data
are of Ref.~\cite{Eberlein-08} (digitized from Ref.~\cite{Despoja-13}).
Inset shows the spectrum calculated using the loss function of Eq.~(\ref{bulk}) (neglect of the elastic scattering).
}
\end{figure}

On the other hand, in Fig.~\ref{Eberlein} we plot results corresponding to the celebrated 
EELS in TEM experiment of Ref.~\cite{Eberlein-08}. Analysis of the geometry of this experiment  shows
that $|\Delta \pv_\| |\approx 3.08$ \AA$^{-1}$, which is outside the first Brillouin zone.
The calculated spectrum ignoring the influence of the elastic scattering 
[obtained by Eq.~(\ref{bulk}) for the energy-loss function] is plotted in the inset of Fig.~\ref{Eberlein},
and it clearly bears no resemblance to the experimental spectrum.
This can be understood considering that, if the elastic channel is switched off, 
the whole (huge) momentum $\Delta \pv_\|$ must be absorbed by the electronic sub-system,
resulting in plasmons and single-particle excitations in higher bands. Although these processes, in fact, take place,
their contribution to the spectrum is negligible when the elastic scattering is taken into account,
which restores a reasonable agreement with experiment, as can be seen in the main panel of Fig.~\ref{Eberlein}. 
The elastic channel accepts the reciprocal-lattice-vector part of the momentum, the rest absorbed by the electronic sub-system.
We emphasize that the above being a reasoning in physical terms,
our Eq.~(\ref{cross2D}) includes all the processes in question, 
producing correct results automatically, without any by-hand manipulations (cf. Ref.~\cite{Despoja-13}, where the
reciprocal lattice vector is subtracted implicitly).

As noted above, this work uses the LDA to the DFT for the ground-state and RPA for the dynamic response calculations, respectively.
This is done consciously for the sake of simplicity and considering that our goal is highlighting the coupling between the inelastic and elastic processes, rather than studying the many-body effects in Q2D materials,
the latter problem having been addressed in a large body of literature  
(see, e.g., Ref~\cite{Siegel-11} and references therein). 
The basis of our approach, Eq.~(\ref{cross2D}), remains, however, valid at any level of (TD)DFT, allowing inclusion of the many-body effects in the framework of this theory. 
At the same time, it must be  noted  that the super-cell method might encounter difficulties in the presence of the long-range exchange and correlations
\cite{Dobson-16},
in which case the `native' approaches \cite{Trevisanutto-16} (considering a single Q2D crystal from the very beginning)  will be necessary.
The well known shortcomings of LDA and RPA \cite{Giuliani&Vignale} are the likely source of the remaining discrepancies between
our calculations and  the experiment in the energy-range of the $\pi+\sigma$ plasmon (see Appendix \ref{AD} for further discussion).

In conclusions, 
we have put forth a theory of electron energy-loss spectroscopy of quasi-2D crystals in the framework of the
quantum-mechanical scattering of a probe electron at a many-electron system, 
accompanied by the elastic scattering at the crystalline
potential. A strong coupling between the inelastic and elastic channels
has been found in graphene, in the incident  energy range
characteristic to the high-resolution EELS ($\sim 10-100$ eV).
This has been shown to be a result of the probe electron interaction with the elastic {\it scattering resonances} 
during its energy-transfer to the electronic sub-system of the target.
In particular, the excitation of the $\pi+\sigma$ plasmon in graphene is 
dramatically affected by the scattering in the elastic channel.
These theoretical findings constitute a strong motivation for conducting  HREELS experiments on free-standing Q2D crystals.
For EELS in the transmission electron microscope ($\sim 40 - 120$ keV), our theory  provides a mechanism 
of the  absorbed momenta distribution between the inelastic scattering 
and the diffraction at the lattice of the Q2D crystal.
By this, a reasonable agreement between the theory and the experimental EELS in TEM has been observed for graphene.

We, finally, argue that, overcoming the limitations of the traditional energy-loss functions formalism,
our approach can be expected to replace it as a standard theoretical tool in the EELS of quasi-2D materials.

Authors are grateful to Dr.~Ming-Wen Chu for providing   digital
EELS experimental data of Ref.~\cite{Liou-15}.
V.U.N. acknowledges support from the Ministry of Science and Technology, Taiwan
(Grant  Nos. 105--2112--M--001--010 and 106--2112--M--001--021). 
This work was supported by the Spanish Ministry of Economy and Competitiveness
MINECO (Project No. FIS2016-76617-P).  


%

\appendix

\onecolumngrid

\section{Connection to the dipole-scattering regime}
\label{AA}

Here we, for simplicity, consider the flat in-plane potential and, therefore, we can rewrite Eq.~(\ref{cross2D}) with all the reciprocal lattice vectors
equated to zero
\begin{equation}
\frac{1}{A} \frac{d \sigma}{d \omega d \Omega} (\pv'\leftarrow \pv) = 
-\frac{64 \pi^5 p'}{p q^2} {\rm Im} 
\int \chi(z_1,z_1',\qv,\omega) 
e^{-q |z_1-z|} e^{- q |z_1'-z'|}   
 a^{+*}_{\pv}(z) a^-_{\pv'}(z)
a^+_{\pv}(z') a^{-*}_{\pv'}(z')  
 d z d z' d z_1 d z_1'.
 \label{cG0}
\end{equation}

The dipole-scattering regime is the one when the target is excited through the long-range Coulomb interaction with a probe,
without the probe charge entering the electron-density of the target. Assuming that the electron-density of the target
and the incident and reflected probe are separated by the $z=0$ plane, we can consider that $\chi(z_1,z_1',\qv,\omega) $ is
nonzero only when both $z_1$ and $z_1'$ are negative, but $a^{\pm}_{\pv}(z)$ are nonzero only if $z$ is positive.
Then Eq.~(\ref{cG0}) can be rewritten as
\begin{equation}
\frac{1}{A} \frac{d \sigma}{d \omega d \Omega} (\pv'\leftarrow \pv) = 
-\frac{64 \pi^5 p'}{p q^2} {\rm Im} 
\int \chi(z_1,z_1',\qv,\omega) 
e^{-q (z-z_1) }e^{- q (z'-z_1')}   
 a^{+*}_{\pv}(z) a^-_{\pv'}(z)
a^+_{\pv}(z') a^{-*}_{\pv'}(z')  
 d z d z' d z_1 d z_1',
\end{equation}
which further reduces to
\begin{equation}
\frac{1}{A} \frac{d \sigma}{d \omega d \Omega} (\pv'\leftarrow \pv) = 
-\frac{64 \pi^5 p'}{p q^2} {\rm Im} 
\int \chi(z,z',\qv,\omega) e^{ q (z+z')}   d z d z' \times
 \left| \int e^{-q z } a^{+*}_{\pv}(z) a^-_{\pv'}(z) d z \right|^2.
 \label{EE}
\end{equation}
We see that the differential cross-section factorizes in this case into the product of two terms. The first
\begin{equation}
 -{\rm Im} g(\qv,\omega)= -{\rm Im} \int \chi(z,z',\qv,\omega) e^{ q (z+z')}   d z d z' 
\end{equation}
is a characteristic of the target only, and it exactly coincides with the minus imaginary part of the $g$-function \cite{Liebsch,Nazarov-15}.
The second term in Eq.~(\ref{EE}) is a purely kinematic factor, depending on the motion of the probe only.

\section{LEED wave-functions' asymptotic  boundary conditions in vacuum}
\label{AB}

The asymptotic behavior of $\langle \rv|\pv^\pm\rangle$ and, hence, 
that of $a^\pm_{\pv,\Gv}(z)$, follows from Eqs.~(\ref{LS}).
Introducing the notation
\begin{equation}
k_\Gv=\sqrt{p_z^2-\Gv^2-2 \Gv\cdot \pv_\| + i 0_+}, \ {\rm Im}\, k_\Gv>0,
\label{kG}
\end{equation}
we can easily find at $k_\Gv^2 \ge 0$
\begin{equation}
a^\pm_{\pv,\Gv}(z) = \delta_{\Gv \0v}  e^{i p_z z} + \left\{
\begin{array}{ll}
b^\pm_{\pv,\Gv} e^{\pm i k_\Gv  z}, &z\rightarrow \infty, \\
c^\pm_{\pv,\Gv} e^{\mp i k_\Gv  z}, &z\rightarrow -\infty.
\end{array}
\right.
\end{equation}
Otherwise, if $k_\Gv^2 < 0$,
\begin{equation}
a^\pm_{\pv,\Gv}(z) = \left\{
\begin{array}{ll}
b^\pm_{\pv,\Gv} e^{-| k_\Gv|  z}, &z\rightarrow \infty, \\
c^\pm_{\pv,\Gv} e^{|k_\Gv|  z}, &z\rightarrow -\infty.
\end{array}
\right.
\end{equation}

\section{LEED wave-functions in the first Born approximation}
\label{LEED}
\label{AC}

From Eqs.~(\ref{LS}) we can write to the first order in $V_l$
\begin{equation}
|\pv^+\rangle=\left[1+G^0 \left(\frac{p^2}{2 }+ i 0_+ \right) V_l \right] |\pv\rangle.
\label{LS1}
\end{equation}

We use the Fourier-series representation of the potential within the interval $z\in \left[-\frac{D}{2},\frac{D}{2} \right]$,
outside of which it is zero 
\begin{equation}
V_l(\rv)= \Omega_D(z) \sum\limits_{\Gv, g}  V_{\Gv, g} e^{i(\Gv\cdot \rv_\|+g z)},
\label{VO}
\end{equation}
where $g= 2\pi n/D$, $n=0,\pm 1,...$, and
\begin{equation}
\Omega_D(z)=\left\{
\begin{array}{ll}
1, & |z| \le D/2,\\
0, & |z|>D/2.
\end{array}
\right. 
\end{equation}

Expanding (\ref{VO}) into the Fourier integral on $z\in (-\infty,\infty)$, we have
\begin{equation}
V_l(\rv)  =  \int  \frac{d g'}{2\pi i g'} 
\sum\limits_{\Gv,g} \!  V_{\Gv, g} e^{i \Gv \cdot \rv_\|} e^{i (g+g') z} 
\left[ e^{i g' D/2} - e^{-i g' D/2}\right].
\label{V1}
\end{equation}
Substituting Eq.~(\ref{V1}) into (\ref{LS1}) and applying $G^0$ explicitly, we have
\begin{equation}
\langle \rv |\pv^+\rangle= \frac{1}{(2\pi)^{3/2}}
\left\{ e^{i \pv\cdot \rv}- \sum\limits_{\Gv, g}   V_{\Gv, g} e^{i(\Gv+\pv_\|)\cdot \rv_\|}  
\int  \frac{d g'}{\pi i g'} \frac{e^{i (g+p_z) z} \left[e^{i g' (z+D/2)} - e^{i g'(z-D/2)}\right]}{(g'+g+p_z)^2 -(p_z^2-\Gv^2-2 \Gv\cdot \pv_\| + i 0_+)}
\right\}.
\end{equation}
After an explicit integration, we have separately in the three regions
\begin{equation}
\langle \rv |\pv^+\rangle = \frac{1}{(2\pi)^{3/2}}
\left\{ e^{i \pv\cdot \rv}- 2 i \sum\limits_{\Gv, g}   V_{\Gv, g} e^{i(\Gv+\pv_\|)\cdot \rv_\|}  
\frac{e^{i k_\Gv z} e^{i g D/2} \sin [(k_\Gv-p_z) D/2] }{(k_\Gv-g-p_z) k_\Gv} \right\}, \ z>D/2,
\end{equation}
\begin{equation}
\langle \rv |\pv^+\rangle = \frac{1}{(2\pi)^{3/2}}
\left\{ e^{i \pv\cdot \rv}-2 i\sum\limits_{\Gv, g}   V_{\Gv, g} e^{i(\Gv+\pv_\|)\cdot \rv_\|}  
  \frac{ e^{-i k_\Gv z} e^{i g D/2}\sin [(k_\Gv+p_z) D/2] }{(k_\Gv+g+p_z) k_\Gv}
\right\}, \ z<-D/2.
\end{equation}
\begin{equation}
\langle \rv |\pv^+\rangle \! = \!  \frac{1}{(2\pi)^{3/2}} \!
\left\{ e^{i \pv\cdot \rv} \! - \!  \sum\limits_{\Gv, g}   V_{\Gv, g} e^{i(\Gv+\pv_\|)\cdot \rv_\|}  
\! \left[
\frac{2 e^{i (g+p_z) z} }{(g \! + \! p_z)^2 \! - \! k^2_\Gv} \! + \!
 \frac{e^{i k_\Gv z} e^{i (k_\Gv-g-p_z) D/2} }{(k_\Gv-g-p_z) k_\Gv}
\! + \!  \frac{e^{-i k_\Gv z}  e^{i (k_\Gv+g+p_z) D/2 }}{ (k_\Gv+g+p_z) k_\Gv}
\right]
\right\}, \ |z|<D/2,
\end{equation}
where $k_\Gv$ is defined by Eq.~(\ref{kG}).
Therefore, with the use of Eq.~(\ref{apm})
\begin{equation}
a^+_{\pv,\Gv}(z)= \frac{1}{(2\pi)^{3/2}}
\left\{ e^{i p_z z} \delta_{\Gv \0v} - 2 i \sum\limits_{g}   V_{\Gv, g} 
\frac{e^{i k_\Gv z} e^{i g D/2} \sin [(k_\Gv-p_z) D/2] }{(k_\Gv-g-p_z) k_\Gv} \right\}, \ z>D/2,
\end{equation}
\begin{equation}
a^+_{\pv,\Gv}(z) = \frac{1}{(2\pi)^{3/2}}
\left\{ e^{i p_z z} \delta_{\Gv \0v}-2 i\sum\limits_{g}   V_{\Gv, g}  
  \frac{ e^{-i k_\Gv z} e^{i g D/2}\sin [(k_\Gv+p_z) D/2] }{(k_\Gv+g+p_z) k_\Gv}
\right\}, \ z<-D/2.
\end{equation}
\begin{equation}
a^+_{\pv,\Gv}(z) \! = \!  \frac{1}{(2\pi)^{3/2}} \!
\left\{e^{i p_z z} \delta_{\Gv \0v} \! - \!  \sum\limits_{ g}   V_{\Gv, g}  
\! \left[
\frac{2 e^{i (g+p_z) z} }{(g \! + \! p_z)^2 \! - \! k^2_\Gv} \! + \!
 \frac{e^{i k_\Gv z} e^{i (k_\Gv-g-p_z) D/2} }{(k_\Gv-g-p_z) k_\Gv}
\! + \!  \frac{e^{-i k_\Gv z}  e^{i (k_\Gv+g+p_z) D/2 }}{ (k_\Gv+g+p_z) k_\Gv}
\right]
\right\}, \ |z|<D/2.
\end{equation}
Finally,  since $ |\pv^-\rangle=|(-\pv)^+\rangle^*$,  $a^-_{\pv,\Gv}(z)$ are found as
\begin{equation}
 a^-_{\pv,\Gv}(z)=a^+_{-\pv,-\Gv}(z)^*.
\end{equation}

\newpage

\section{Further  comparison with experiment}
\label{AD}

\begin{figure}[h!]
\includegraphics[width=  \columnwidth, clip=true, trim=60 0 10 0]{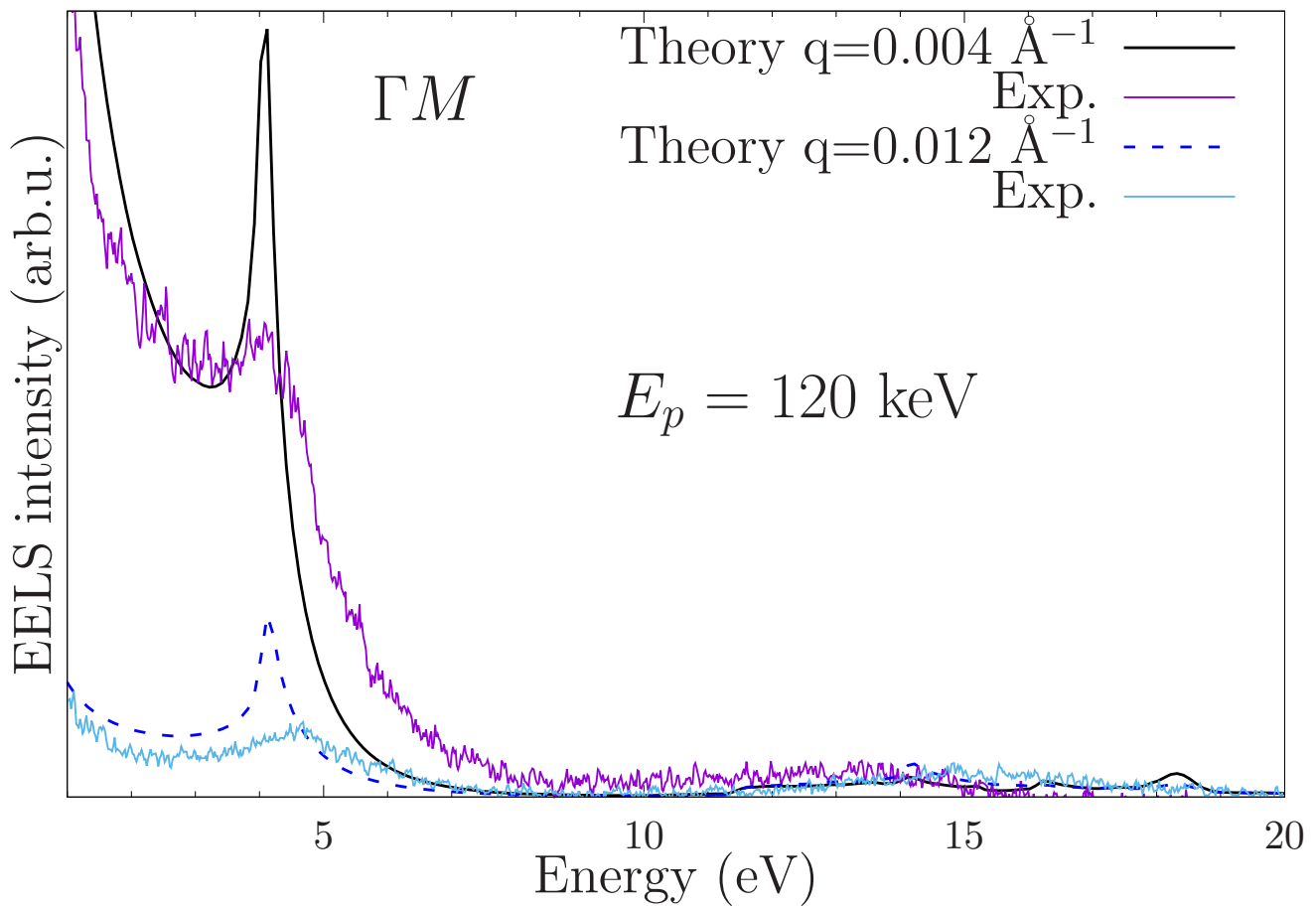}
\caption{\label{Liou}
Calculated EEL spectrum of graphene
(smooth solid lines) and the experimental EELS in TEM (noisy lines). Experimental data are
from Ref.~\cite{Liou-15}. 
}
\end{figure}

Although theoretical  spectra in Figs.~\ref{Wachsmuth} and \ref{Eberlein} are in a qualitative agreement with the experimental EELS in TEM,
two  differences can be noticed. First, in Fig.~\ref{Wachsmuth} in the energy range below the $\pi$-plasmon peak, the intensities of the experimental spectra are greater than those of the theoretical ones. 
We attribute this to the finite momentum resolution ($\Delta q=0.1$\AA$^{-1}$) in the experiment \cite{Wachsmuth-13}.
Indeed, the growth of the intensity with the decreasing energy below the $\pi$-peak 
is characteristic for smaller wave-vectors \cite{Wachsmuth-13} (see also Fig.~\ref{Liou}).
Because of the contribution from the smaller $q$-s than the nominal one, this
leads to the discrepancy between the theory and experiment in this energy range.

Secondly, the experimental $\pi+\sigma$ plasmon is well reproduced by our calculations except for the amplitude around the maxima.
This feature is persistent with respect to the change of $q$ and, therefore, is likely to be related to the shortcomings of the LDA and RPA used in the calculations. 

Finally, in Fig.~\ref{Liou} we present  theoretical spectra in comparison with the recent experimental EELS  in TEM at very small wave-vectors \cite{Liou-15}.
At so small $q$-s, the theoretical $\pi$-peak in pristine graphene 
is almost dispersionless in LDA and RPA, which is due to the overlapping of the plasmon with the inter-band transitions \cite{Nazarov-15}.
This finds itself in contrast with the experimental behavior  \cite{Liou-15}. The inclusion of the static exchange and correlations
beyond the LDA and going beyond RPA by accounting for the {\it dynamic} exchange and correlation with the use of $f_{xc}$
in Eq.~(\ref{TDDFT})
of a sufficient degree of sophistication, may be the way to resolve this discrepancy.

\end{document}